\documentclass[5p, preprint]{elsarticle}

\usepackage{physics}
\usepackage{siunitx}

\usepackage{hyperref}

\usepackage{soul} 
\usepackage{tikz} 

\journal{Carbon}

\begin{document}

\begin{frontmatter}

\title{Numerical Simulations of Pressure Induced sp$^2$-sp$^3$ Transitions in Defect Carbon Nanotubes}

\author[a1]{Xolani Maphisa\corref{cor1}}\ead{maphisaxolani13@gmail.com}
\author[a1]{Robert Warmbier}\ead{rwarmbier@uj.ac.za}
\cortext[cor1]{Corresponding author}
\address[a1]{Department of Physics, University of Johannesburg, P.O. Box 524, 2006 Auckland Park, South Africa}

\begin{abstract}
The combination of pressure and vacancy defects are investigated to find the ideal conditions that would create meaningful sp$^3$ interlinking without causing severe damage to the single-walled carbon nanotubes (SWCNTs). Naturally occurring defects fail to induce interlinking. The introduction of vacancy type defects reduces the collapse pressure of the SWCNTs. The combination of vacancy defects, high temperature and low pressure induces sp$^3$ without causing severe damage to the nanotubes. Structural changes caused before and after pressurization of the nanotubes were analyzed using Raman spectroscopy.
\end{abstract}


\end{frontmatter}


\section{Introduction}
Carbon nanotube (CNTs) have received a great deal of interest given their outstanding combination of properties and prospective applications \cite{ajayan2001applications}. In particular the combination of high strength and flexibility as well excellent thermal and electrical conductivity come to mind. Individual single wall carbon nanotubes have shown a Young's modulus of about \SI{1000}{GPa} and tensile strength above \SI{100}{GPa} \cite{salvetat1999elastic,lu1997elastic}.\\
Materials have different properties than constituents, which can also be size-dependent in the case of nano-structured allotropes. In the cases of nanotubes this difference is significant as many of the desirable properties are severely diminished \cite{kis2004reinforcement}. Adjacent tubes are held together by van der Waals forces only, so that tubes can slide past each other easily\cite{ajayan2001applications}. The lack of strong interaction between the tubes also limits the overall thermal and electrical conductivity. Similar effects are reported with double walled carbon nanotubes (DWCNTs), where the poor linkage exists not only between the outer adjacent tubes, but also between inner and outer tubes.\\
The introduction of covalent interlinks between adjacent CNTs can improve their mechanical and electrical properties and open up more applications \cite{xia2007enhancing}. Prevalent strategies used to promote sp$^3$ interlinks in CNTs involve ion or electron irradiation and pressurization.
Irradiation has been found to generate interlinks, it inevitably also damages the nanotubes\cite{o2013improved, peng2008measurements, gupta2007changes, ritter2006radiation}. At higher dosages this leads to a significant degradation of the nanotubes and a loss of mechanical stability. High pressure can also be used to promote interlinks at the high curvature points resulting from the deformation. It has been observed that sp$^3$ interlinks induced at low pressure are often reversible when the structure is recovered to ambient conditions \cite{loa2003raman}. In contrast going to high pressure has been shown to cause irreversible structural damage \cite{loa2003raman, jensen2020toward, sakurai2011pressure} as can be expected.\\ 
As any method able to promote interlinking is likely to also cause some damage to the material, a difficult balance needs to be struck between those competing effects. Combining both irradiation and pressure might be avenue to obtain better results. A combination of low fluence ion (B$^+$) irradiation and pressurization to generate sp$^3$ interlinks in DWCNTs has produced some promising results\cite{hearne2021effects}. The low radiation dosage would create only a small number of defects in the sample, avoiding significant damage. The low defect density may not immediately generate a significant amount of interlinks, they might however promote increased interlinking when the sample is exposed to pressure. This combination of irradiation and pressure is expected to lead to permanent interlinks using low ion irradiation and pressure levels only, thus avoiding most of the detrimental effects of these approaches.\\
In this paper, we build upon this approach and use numerical tools to investigate how the combination of defects and pressure induces interlinks. For this we employ Molecular Dynamics simulations of defect nanotubes under pressure. Resultant interlinks are further characterized using Density Functional Theory. To further characterize the induced changes we compute the corresponding Raman spectra.

\section{Methodology}
The same set-up of the simulation cells were used for both Molecular Dynamics (MD) and Density Functional Theory (DFT) simulations. The computational cost of DFT require simulations cells of no more than a few hundred atoms. Furthermore, interlinking can be simulated easiest using aligned nanotubes. We therefore used two parallel CNTs per unit cell in a close pack configuration, which is the smallest possible configuration to observe adequate deformation under pressure. (5,5) single wall CNTs with diameter of about \SI{7}{\angstrom} were chosen as they are small enough to contain few atoms around the circumference, but are large enough to be commonly found in real samples. With a length of about \SI{20}{\angstrom} each unit cell had a total of 320 carbon atoms except for cases where vacancy type defects were created. Periodic boundary conditions were used in all directions. At creation the spacing between all CNTs was set to \SI{3.4}{\angstrom}, which is roughly the separation between graphene sheets. Simulation structures were created and modified using the Atomic Simulation Environment (ASE) software \cite{larsen2017atomic}.\\
All molecular dynamics simulations were performed using the open source Large-scale Atomic/Molecular Massively Parallel simulator (LAMMPS) software \cite{plimpton1995fast}. The ReaxFF\textsubscript{C-2013} force field was used as implemented in LAMMPS, as it is known to describe sp$^2$ as well as sp$^3$ bonds accurately. MD simulations were performed using a time step of \SI{0.2}{fs}, which tests showed to be sufficient. The Nos\'e-Hoover thermostat and barostat were used for temperature and pressure control. A constant temperature of \SI{100}{K} was set unless stated otherwise and isotropic pressure was applied. The pressure was increased in intervals of \SI{5}{GPa} at a rate of \SI{50}{GPa} \si{nm^{-1}}, unless stated otherwise, until total collapse was observed. At the end of each interval the system was allowed to equilibrate for \SI{10}{ps} before proceeding.\\
The DFT calculation were performed using the GPAW software \cite{mortensen2005real, enkovaara2010electronic}, using an LCAO DZP basis as shipped with GPAW. The PBE xc-functional \cite{perdew_generalized_1996, perdew_generalized_1997} was used with $\Gamma$-point only to describe the Brillouin zone. The Raman spectrum was computed using the 3rd order perturbation theory as implemented recently in GPAW. Structure for Raman simulations were taken from MD, but had to be relaxed with fixed lattice vectors to obtain a force-free reference geometry.\\
Visualizations of atomic structures were created using ASE\cite{larsen2017atomic}, Ovito \cite{stukowski2009visualization} and VESTA\cite{vesta}.

\section{Pressure Simulations}
\subsection{Collapse Pressure}
\begin{figure}[tb]
	\centering
	\includegraphics[width=0.45\textwidth]{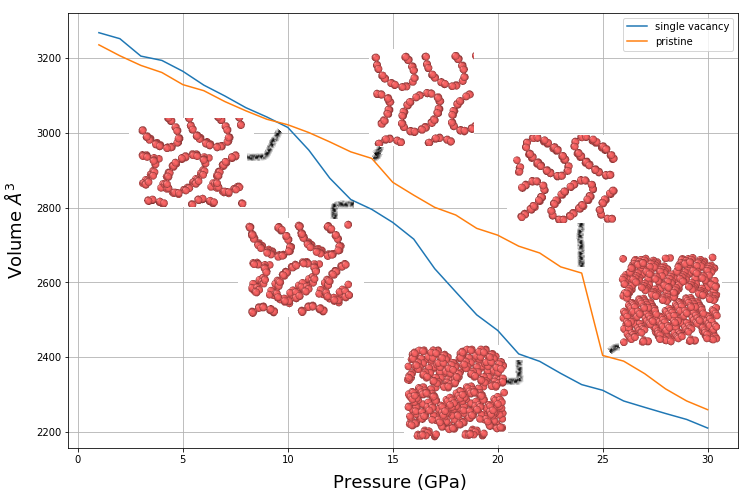}
	\caption{Relationship between volume and pressure applied for (5,5) pristine and single vacancy defect CNT bundles. Insets show the structural changes at selected points.}
	\label{fig:volume_vs_pressure}
\end{figure}

In order to obtain a baseline for the different pressure regimes for the system at hand a pressure series was recorded in steps of \SI{1}{GPa}. As pressure increases a system undergoes consecutive geometrical transitions, which are marked by abrupt change in volume. As CNTs are low density structures which collapse under pressure, these changes are quite prominent, as shown in Fig.~\ref{fig:volume_vs_pressure}. The insets show the corresponding structural changes at selected points. In addition to the pristine cell, a system with 2 single vacancy defects is shown as well.\\
The pristine nanotube bundle behaves somewhat differently from defect nanotubes under pressure despite the relatively low defect content of \SI{0.6}{\percent}. As can be expected. The volume of the pristine and defect systems are almost identical before pressurization, with small differences caused by numerical fluctuations. The tubes change from circular to elliptical shape in the first phase transition, which occurs at \SI{10}{GPa} for the defect system and \SI{15}{GPa} for the pristine system. The pristine system thereafter shows again a linear compression characterized by a reduction in CNT-CNT distance and increase in CNT eccentricity before a rapid total collapse at \SI{25}{GPa}. The defect CNT system shows a qualitatively different behavior in the second phase. In addition to the effects discussed, it also shows signs of gradual collapse, with the final collapse at \SI{21}{GPa} being much less pronounced as a result. After the total collapse, happening at different pressures but same volumes, the systems behave identically, as notions of the original structures are barely existing anymore.\\
For pristine nanotubes, interlinking can only be facilitated by re-hybridization at high curvature points as created for high pressure. This process is of course in competition with the total, irreversible collapse of the nanotubes. As can be seen from Fig.~\ref{fig:volume_vs_pressure}, even a small admixture of defects moves the creation of high curvature points to lower pressure and away from the collapse edge.

\subsection{Pressure Behavior of Pristine Nanotubes}
\begin{figure}[tb] 
	\begin{center}
		\includegraphics[width=0.4\textwidth]{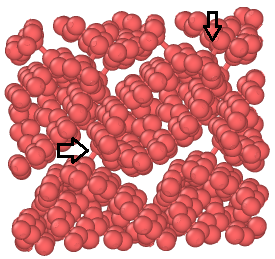} 
	\end{center}
	\caption{Snapshot of a defect-free CNT bundle during pressure ramp \SI{35}{GPa}-\SI{40}{GPa} at \SI{100}{K}.}
	\label{fig:sidewalls}
\end{figure}
Looking closer at MD simulations at high pressure shows that around the total collapse pressure potential CNT-CNT interlinks are formed. An example is shown in Fig.~\ref{fig:sidewalls}. When the CNTs collapse, the sp$^2$ bonds in the nanotubes re-hybridize, creating sp hybridized orbitals at the high curvature points, which can act as a reactive site. In certain cases the interlinks observed in the collapsed structure remain when the pressure is removed. This is however primarily the case when collapse was not completely reversible.\\
\begin{figure}[tb]
	\centering
	\includegraphics[width=0.40\textwidth]{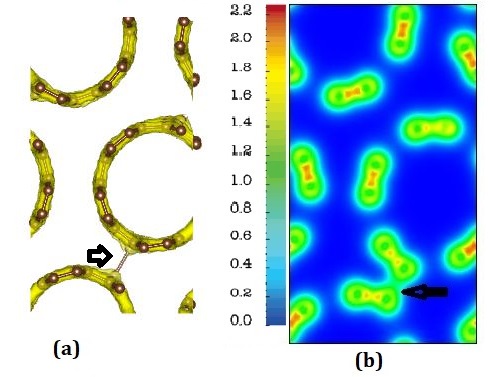}
	\caption{Electron density plots of pristine CNT bundle obtained after pressure is removed from \SI{40}{GPa} at \SI{100}{K}.}
	\label{fig:100K_ed}
\end{figure}
An example of this can bee seen in Fig.~\ref{fig:100K_ed}, which shows electron density plots for the final pristine nanotubes at \SI{100}{K} after the pressure was reversed at \SI{40}{GPa}. In this case the structural changes under compression were reversible, an almost perfect circular cross-section is recovered. As potential interlink is indicated in Fig.~\ref{fig:100K_ed}~(a) by VESTA. However neither electron density plot in Fig.~\ref{fig:100K_ed} suggests any kind of bonding. Such artificial 'bonds' drawn by visualization software are common as those work by comparing inter-atomic distances to covalent and/or ionic radii only. In articles using MD simulations it is at times, assumed by the authors that the bonds drawn by the software are real, however this is not always the case, as was shown above. An example relevant to this work is a paper by Sakurai \textit{et al.} \cite{sakurai2011pressure}, where the authors concluded that the bonds drawn by the software were conclusive proof that interlinks are formed. This shows the necessity of verifying the interlinks by other means.\\
\begin{figure}[tb]
	\centering
	\includegraphics[width=0.3\textwidth]{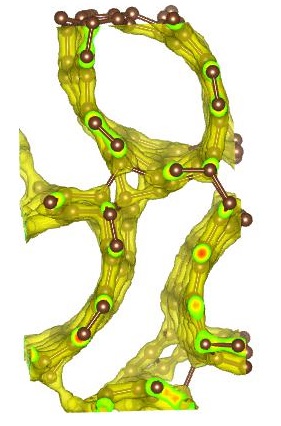}
	\caption{Electron density iso-surface plot of pristine CNT bundle obtained after pressure is removed from \SI{30}{GPa} at \SI{500}{K}.}
	\label{fig:iso_500K}
\end{figure}
Another example is shown in Fig.~\ref{fig:iso_500K} the a bundle was recovered to ambient pressure after it suffered irreversible changes. In this case a number of CNT-CNT interlinks are retained, but the mechanical properties of this sample might be compromised by the damage.

\subsection{Vacancy Reconstruction}\label{sec:vac_recon}
In numerical simulations defects are typically created \textit{ad-hoc} by manually adding, removing or moving atoms in the simulation cell. Care must be given to properly relax the structures obtained in this way to avoid nonphysical results and MD simulations might be required to capture any reconstruction or healing of defects.\\
\begin{figure}[tb]
	\centering
	\includegraphics[width=0.47\textwidth]{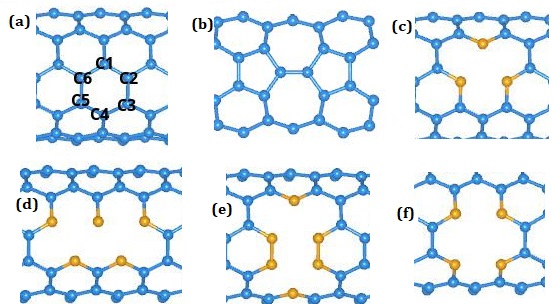}
	\caption{Structures of (a) pristine CNT and (b)-(f) different vacancy defects. (b) Stone-Wales defect, created by rotating C5 and C6 at a \SI{90}{\degree} angle. (c) Single vacancy defect, created by removing C4. (d) Horizontal di-vacancy, created by removing atoms C2 and C6. (e) Vertical-1 di-vacancy, created by removing atoms C1 and C4. (f) Vertical-2 di-vacancy created by removing atoms C2 and C3.}
	\label{fig:defect_types}
\end{figure}
There are many kinds of defects that are induced during irradiation. Here single and double vacancy defects displayed in Fig.~\ref{fig:defect_types}. The Stone-Wales defect (5-7-7-5) are not necessarily created during irradiation, they are naturally occurring defects which can be found in most sp$^2$ carbon materials \cite{ma2009stone}. This defect is not a vacancy defect and it behaved like that pristine system in the MD simulations. It is therefore not further discussed. The vacancy defect structures described here had a total of 318 carbon atoms per unit cell.\\
\begin{figure}[tb] 
	\centering
	\includegraphics[scale=0.40]{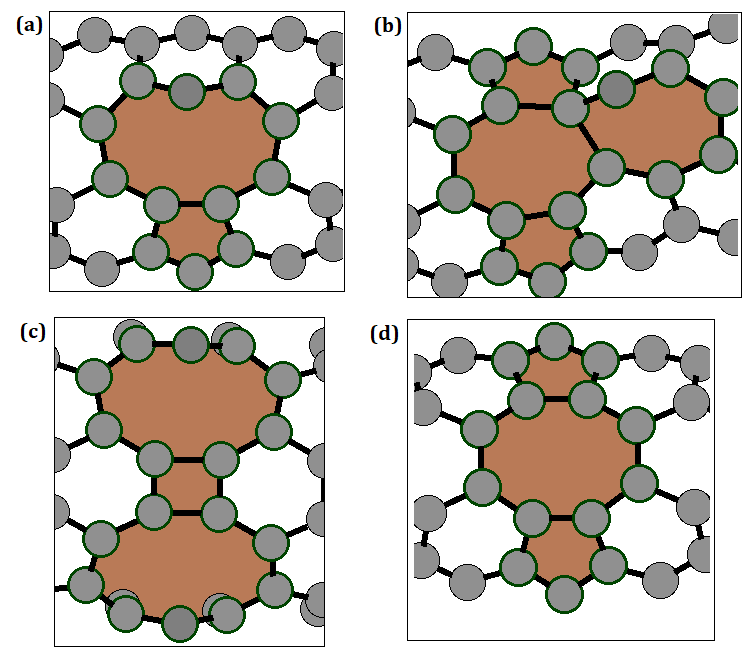}
	\caption{Reconstructed (a) single vacancy defect, (b) horizontal di-vacancy defect type-1, (c) vertical-1 di-vacancy defect and (d) vertical-2 di-vacancy defect.}
	\label{fig:vacancy_reconstruction}
\end{figure}
After allowing the vacancy defects to heal during a dynamic simulation they form more complex topological defects. Fig.~\ref{fig:vacancy_reconstruction} shows the topological defects that resulted from the self healing of the carbon network under pressure. The single vacancy reconstructed to form a pentagon and a nine membered ring (5-9). The single vacancy reconstruction under pressure shown in Fig.~\ref{fig:vacancy_reconstruction}~(a) is similar to that found by Lee \textit{et al.} \cite{lee2009reconstruction}. The vertical-2 di-vacancy, Fig.~\ref{fig:vacancy_reconstruction}~(d), reconstructs to form two pentagons and an octagon, a topological defect known as the 5-8-5 defect, which has also been observed in graphene \cite{amorim2007divacancies}. The horizontal di-vacancy reconstructs to form two pentagons and two seven membered rings. This 5-7-7-5 defect is not similar to the 5-7-7-5 Stone-Wales defect. The one seven membered ring shown in the reconstructed horizontal di-vacancy has an unsaturated carbon atom, making this topological defect less stable than the Stone-Wales topological defect. The vertical-1 defect reconstructs to forms a two nine membered rings and and a square (9-4-9), shown in Fig.~\ref{fig:vacancy_reconstruction}~(c).\\
None of these reconstructed defects showed any improved interlinking in the MD simulations. It should be noted though, that the exact placement of defects in CNT systems heavily influence the behavior under pressure. This makes generalized conclusions rather difficult.

\subsection{Increased Defect Density}
In the previous Section the defects could be viewed as isolated. In order to simulate the effect to interacting defects the number of defect sites was doubled, resulting in 316 carbon atoms per unit cell.\\
With increased defect density it was observed that the final structures had CNT-CNT interlinks after the pressure was removed. Earlier it was observed that the introduction of vacancy defects reduces the collapse pressure. This plays a significant role in the formation of sp$^3$ interlinks. Fig.~\ref{fig:volume_vs_pressure} shows that the volume decreases faster as a function of pressure if vacancy defects are present. As a result the inter-tube distances reduce much quicker, producing a more favorable environment to form interlinks at lower pressure.\\
\begin{figure}[tb]
	\centering
	\includegraphics[width=0.45\textwidth]{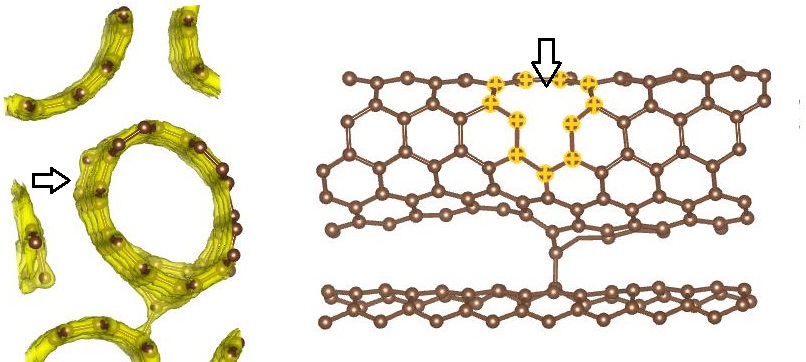}
	\caption{(5,5) CNT bundle with 2 vertical-2 di-vacancies before pressurization. Structures shown were obtained after pressurization: (Left) 3D electron density plot. (Right) Structure without an iso-surface, where the reconstructed vertical-2 defect is shown.}
	\label{fig:Multilple vacancies}
\end{figure}
Fig.~\ref{fig:Multilple vacancies} shows a system which was created with vertical-2 di-vacancies before pressurization. After the pressure cycle up to \SI{30}{GPa} at \SI{100}{K}, the final structure shows an interlink one of the original defects and the other defect did not reconstruct into a topological defect. The difference to the results from Section~\ref{sec:vac_recon} could stem simply from more lucky placement of the defects. It is however plausible to argue, the that higher defect density weakens the nanotube further, in particular allowing for more motion and deformation in the off-axis directions. The different reconstruction of the non-linking defect is in support of the idea, that the two defects directly or indirectly influence each other enough to change their dynamic behavior.

\section{Raman Spectroscopy}
In molecular dynamics simulations the structural transitions under pressure and the pressure induced changes such as deformation of the CNTs or sp$^3$ interlinks can be observed directly. However in experiment this direct observation is not possible. Raman spectroscopy is a non-destructive tool that has been used to study carbon nanotubes under different experimental conditions. It can be used to determine the diameter of CNTs, structural modifications and the the degree of purity and so much more.\\
In the GPAW Raman code all 9 independent polarization directions can be obtained independently and at freely choosable laser wavelength. CNTs are often used in powder form in experiments, which prevents the selection of specific polarization directions. Therefore the Raman spectra shown here are averaged over all directions. Results for a laser wavelength of \SI{532}{nm} are shown here, a common Raman laser wavelength, which produced fairly little noise in the simulations compared to other wavelengths. It should be noted though, that DFT calculations are not very good at capturing conduction band energies. The numerical laser wavelength can therefore not be equated with a specific experimental laser. Lastly, numerical line positions will always deviate somewhat from experimental values due to the numerous approximations involved. That does whatever not devalidate the trends observed.

\subsection{Raman Spectrum of Pristine Nanotubes at High Pressure}
\begin{figure}[tb] 
	\centering
	\includegraphics[width=0.45\textwidth]{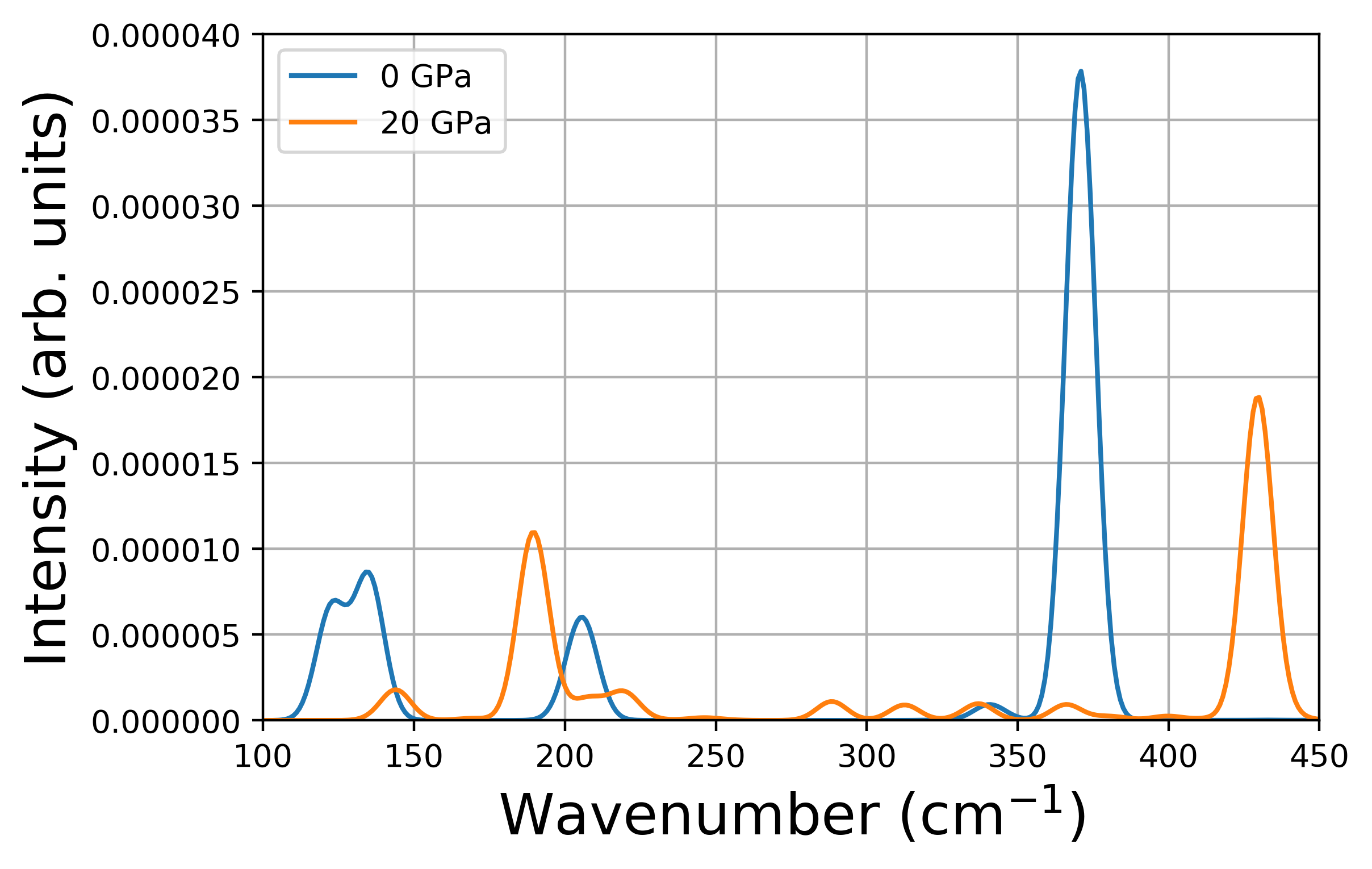}\\
    \includegraphics[width=0.45\textwidth]{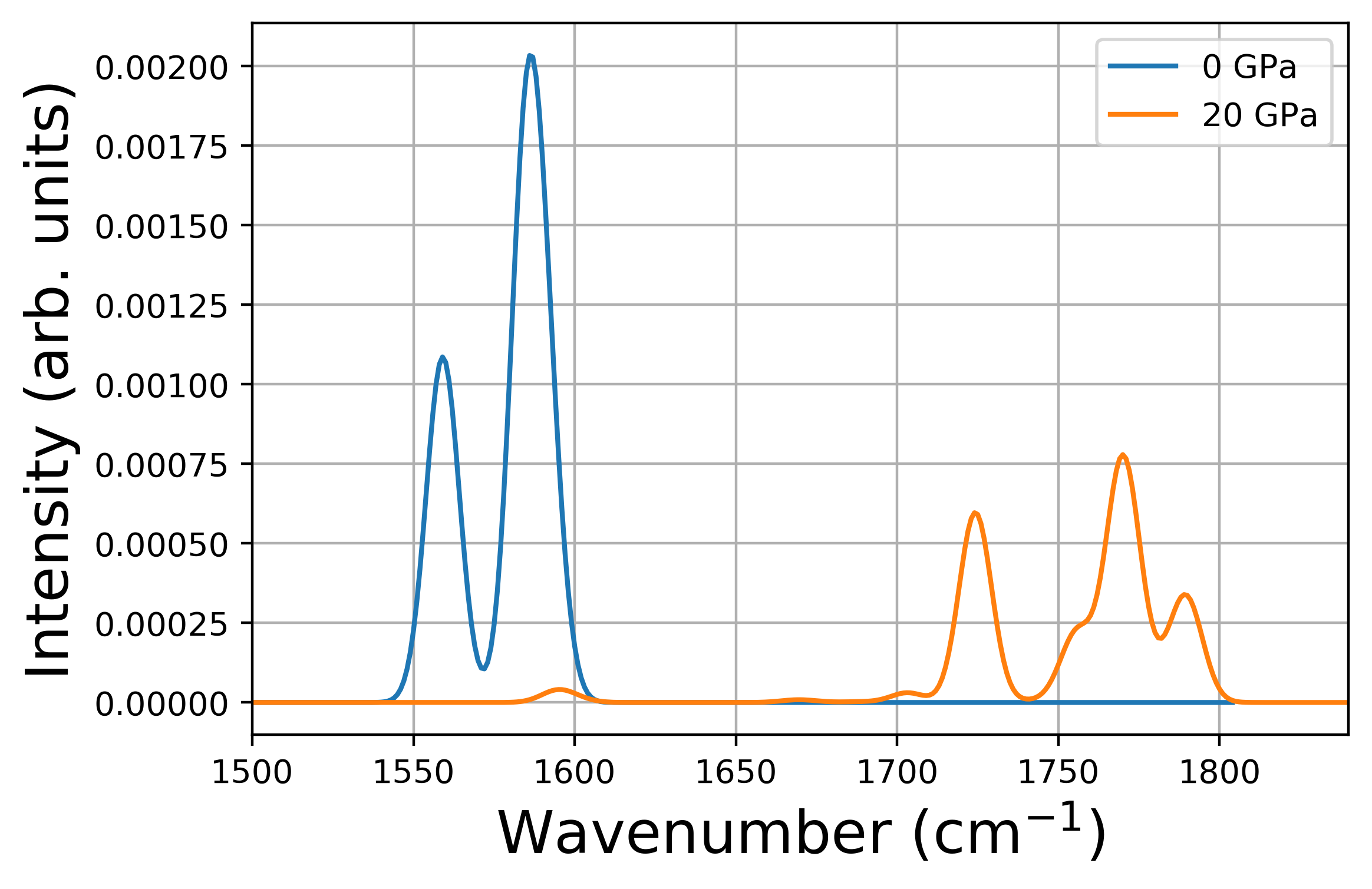}
	\caption{Comparison of the Raman spectrum at \SI{532}{nm} laser wavelength for the (5,5) pristine nanotube bundle at \SI{0}{GPa} and \SI{20}{GPa}.}
	\label{fig:raman_pressure_pristine}
\end{figure}
As shown in Fig.~\ref{fig:volume_vs_pressure}, pristine (5,5) nanotube bundles undergo an ovalization geometric change before the final collapse. At a pressure of \SI{20}{GPa} this is already well pronounced with the CNTs having almost a race-track shape. \\
A comparison between the \SI{0}{GPa} and \SI{20}{GPa} Raman spectrum is shown in Fig.~\ref{fig:raman_pressure_pristine}. All three primary effects of pressure are evident in the spectrum. The deformation caused by the pressure reduces the symmetry of the nanotubes, which makes more modes Raman. The lowered symmetry also causes some modes to split, particularly the ones exhibiting motion in the off-axis directions. Additionally, features are shifted and/or changed in intensity. Several experimental studies have shown that the radial breathing mode (RBM) pressure response is such that the position shifts at a certain rate as pressure increases, whilst other investigations have revealed that the intensity of the RBM drops with increasing pressure \cite{lebedkin2006raman, peters2000structural, merlen2005resonant, venkateswaran2001high, thomsen1999raman}.\\
\begin{figure}[tb] 
	\centering
	\includegraphics[width=0.18\textwidth]{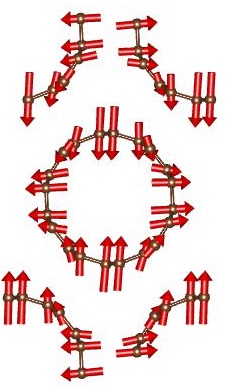}
	\includegraphics[width=0.18\textwidth]{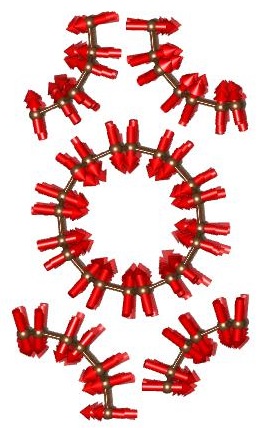}\\
	\includegraphics[width=0.4\textwidth]{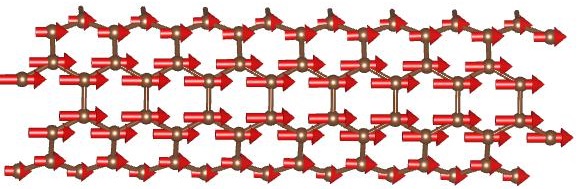}
	\caption{Vibrational motion of atoms giving rise to the modes observed in the low frequency region of the Raman spectrum shown at ambient pressure. (Top left) E$_{2g}$ symmetry mode with peaks at \SI{122.5}{cm^{-1}} and \SI{135}{cm^{-1}}. (Top right) A$_{1g}$ symmetry mode (RBM) at \SI{371}{cm^{-1}}. (Bottom) E$_{1g}$ symmetry mode \SI{205}{cm^{-1}}, side view of the CNT.}
	\label{fig:low_frequency_modes}
\end{figure}
In the low frequency region, two of the three peaks detected before pressurization had shifted to higher frequencies. The E$_{2g}$ symmetry mode had a off-diagonal component at \SI{122}{cm^{-1}} and a diagonal component at \SI{135}{cm^{-1}}. For displacement patterns see Fig.~\ref{fig:low_frequency_modes}. Under pressure its two components merge and shift to \SI{144}{cm^{-1}} with a significant decrease in intensity. This mode behaves similar to the RBM and the observed decrease in intensity could be attributed to the observed structural transition. Unlike the other two modes found in the low frequency region, the E$_{1g}$ symmetry mode does not display any breathing like behavior. As a result, the behavior of this mode under pressure is different. The mode is down-shifted by \SI{15}{cm^{-1}} from its previous position (\SI{205}{cm^{-1}}), additionally the intensity of the mode increases a bit. The motion of this mode is an in-plane change of bond angle, which is not hampered by pressure or the ovalization. The A$_{1g}$ radial mode is up-shifted from \SI{371}{cm^{-1}} to \SI{424}{cm^{-1}} at high pressure with a reduction in intensity. This is in agreement several theoretical and experimental observations \cite{merlen2005resonant, venkateswaran2001high, thomsen1999raman, elliott2004collapse}. Like the E$_{2g}$ mode, the attenuation of the RBM's intensity can be interpreted as a signature of a geometrical transition.\\
In the high frequency region the position of the G-band (G$^-$ and G$^+$) gets blue-shifted with increasing pressure. This is related to hardening of the carbon-carbon interactions. Assuming the peak varies linearly with pressure as predicted by \cite{thomsen1999raman} for SWCNTs, the G$^+$ shifts at a rate of \SI{9.2}{cm^{-1}\per GPa}. However some studies have predicted a multi-step behavior of this mode \cite{yao2008raman}. Apart from blue-shifting and lowered mode intensity, the G$^-$ component showed no additional changes. The G$^+$ component, on the other hand, showed considerable modifications. Multiple components of this mode were detected; the first is the 'shoulder' at \SI{1750}{cm^{-1}}, the second is the most intense peak at \SI{1770}{cm^{-1}}, and the third is situated at \SI{1790}{cm^{-1}}. 


\subsection{Raman Spectrum of Defect Nanotubes before Pressurization}
\begin{figure}[tb] 
    \centering
     \includegraphics[width=0.45\textwidth]{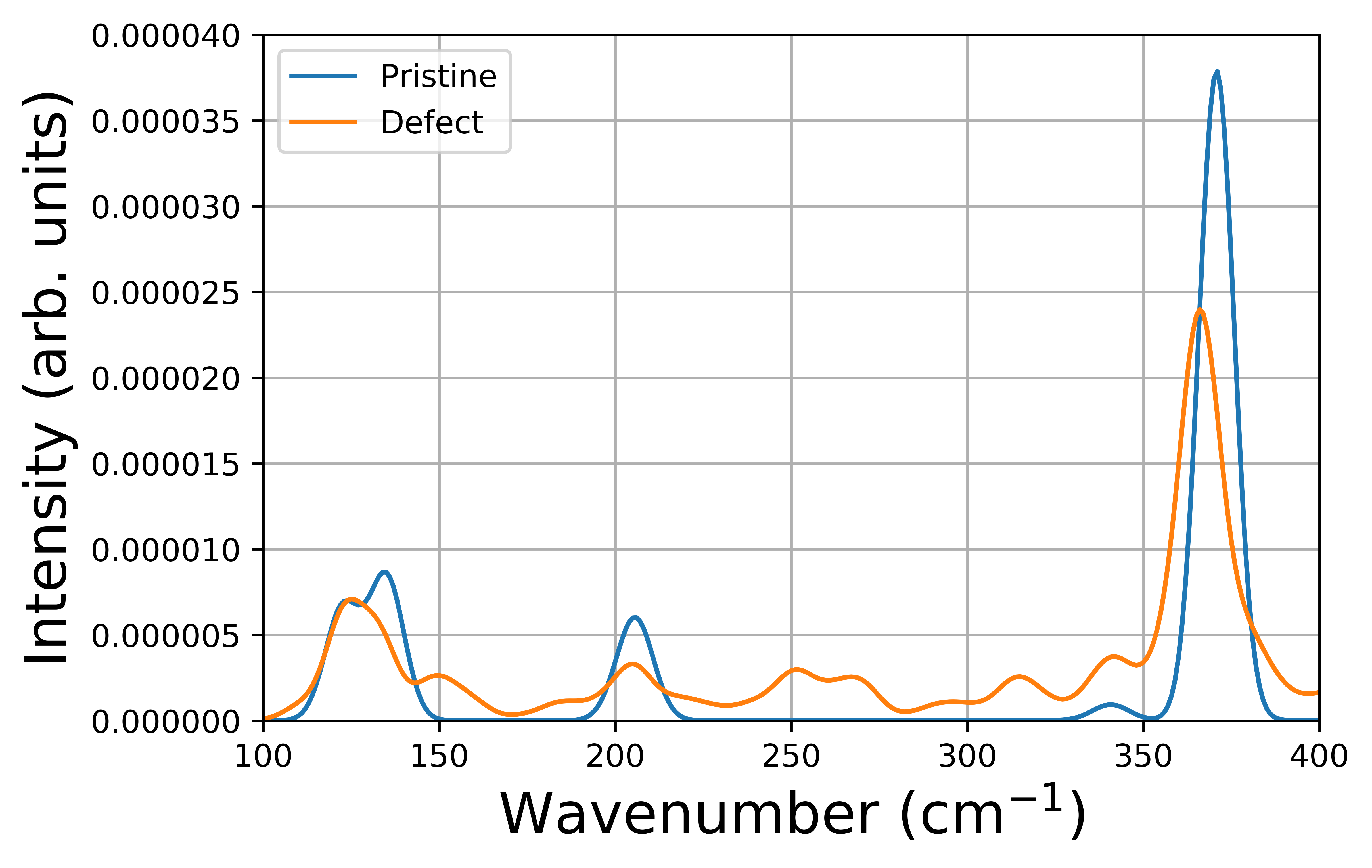}
     \includegraphics[width=0.45\textwidth]{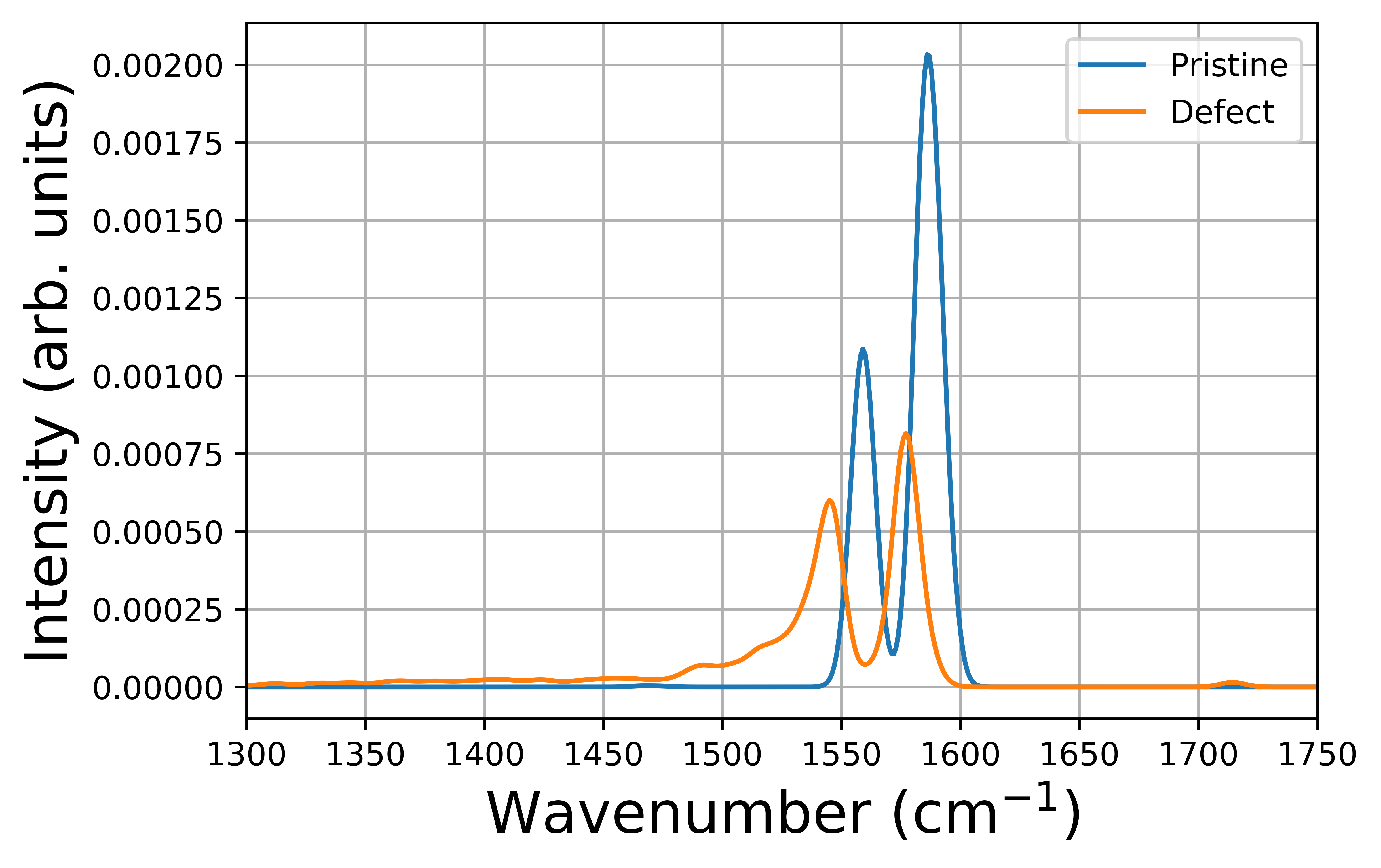}
    \caption{Comparison of the Raman spectrum at \SI{532}{nm} laser wavelength of the (5,5) pristine nanotube bundle and defect nanotube bundle before pressurization.}
    \label{fig:Defect_Spectra}
\end{figure}
The Raman spectrum of single vacancy defect system with with 316 atoms was computed.
The introduction of vacancy defects into the nanotube has only minor influence on the Raman spectrum, as can be seen in Fig.~\ref{fig:Defect_Spectra}. In the low frequency region additional low intensity signals appear, most likely facilitated by the (locally) lowered symmetry caused by the defects. In practise those would most likely be considered noise. The diagonal component of the E$_{2g}$ mode at \SI{135}{cm^{-1}} is right shifted to \SI{151}{cm^{-1}} and decreases significantly in intensity. The introduction of defects causes a change in the vibrational motion of the atoms that give rise to this mode, hence the observed changes in the Raman spectrum.\\
\begin{figure} [tb]
    \centering
     \includegraphics[width=0.45\textwidth]{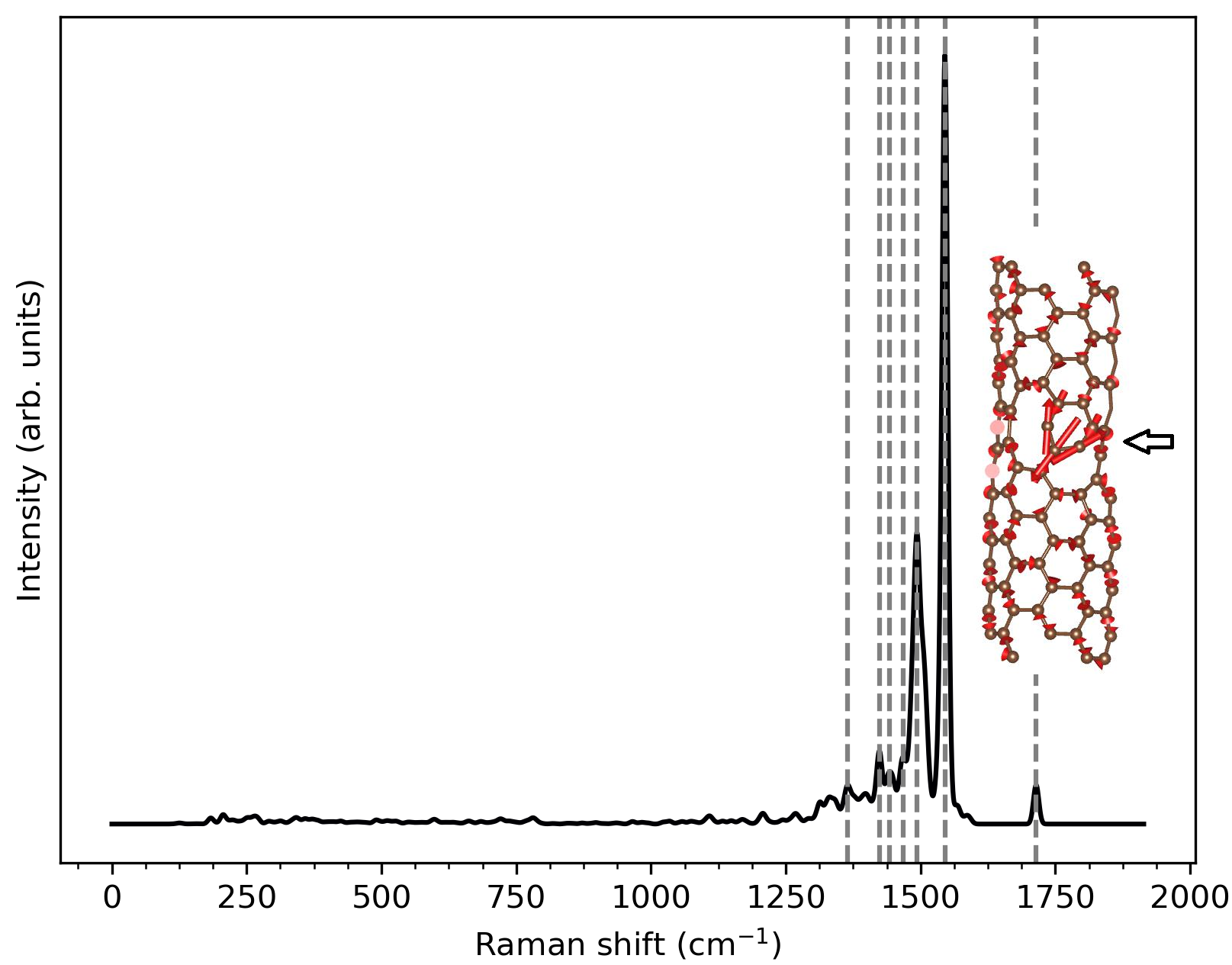}
    \caption{The Raman spectrum of a (5,5) nanotube bundle with single vacancy defects computed at a laser wavelength of \SI{532}{nm} in the $XZ$ polarization direction. The inset illustrates the vibrational motion of atoms that gives rise to the Raman mode at \SI{1714}{cm^{-1}}.}
    \label{fig:Display_D-band}
\end{figure}
The introduction of defects cause the D-band to become Raman active. This mode is not clearly visible in Fig.~\ref{fig:Defect_Spectra} as it occurs only in few polarization directions and with low intensity. However it is more visible in the Raman spectrum computed in the $XZ$ polarization direction shown in Fig.~\ref{fig:Display_D-band}. The D-band is located around \SI{1340}{cm^{-1}} in the displayed Raman spectrum. The G$^-$ component down-shifts to \SI{1545}{cm^{-1}} and the G$^+$ downshifts to \SI{1577}{cm^{-1}}. The down-shifting of the G-band has been observed in irradiated double- and multi-walled carbon nanotubes \cite{aitkaliyeva2014radiation} as a result of weakened the C-C bonds. The Raman spectrum of the nanotube bundle with defects displays another peak at \SI{1714}{cm^{-1}}. This mode is barely visible in the spectrum averaged over all polarization directions in Fig.~\ref{fig:Defect_Spectra}, however, like the D-band it is clearly visible in the $XZ$ polarization direction as shown in Fig.~\ref{fig:Display_D-band}. On the inset, the atomic displacements that induce the mode are displayed, and it can be observed that only the sp bonded atoms in the defect are vibrating. 

\subsection{Raman Spectrum of Defect Nanotubes Recovered to Ambient Pressure}
\begin{figure}[tb]
    \centering
     \includegraphics[width=0.45\textwidth]{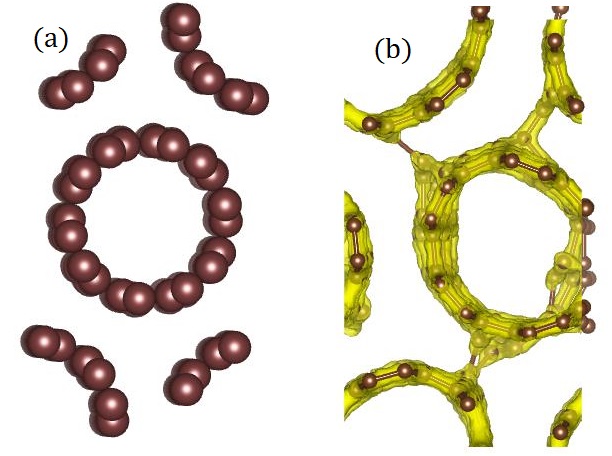}
    \caption{ (a) (5,5) CNT bundle with 4 single vacancy defects before pressurization at \SI{0}{GPa}. (b) 3D electron density plot of the (5,5) CNT bundle at \SI{0}{GPa} after the pressure was reversed from \SI{30}{GPa}.}
    \label{fig:recovered}
\end{figure}
The defect CNTs used in the preceding Section were pressurized up to \SI{30}{GPa} at \SI{100}{K}, after which the structure was allowed to return to ambient pressure. Fig.~\ref{fig:recovered} shows the defect structure before and after pressurization. It can be seen that the final structure has sp$^3$ interlinks between adjacent CNTs.\\  
\begin{figure} [tb]
	\centering
	\includegraphics[width=0.45\textwidth]{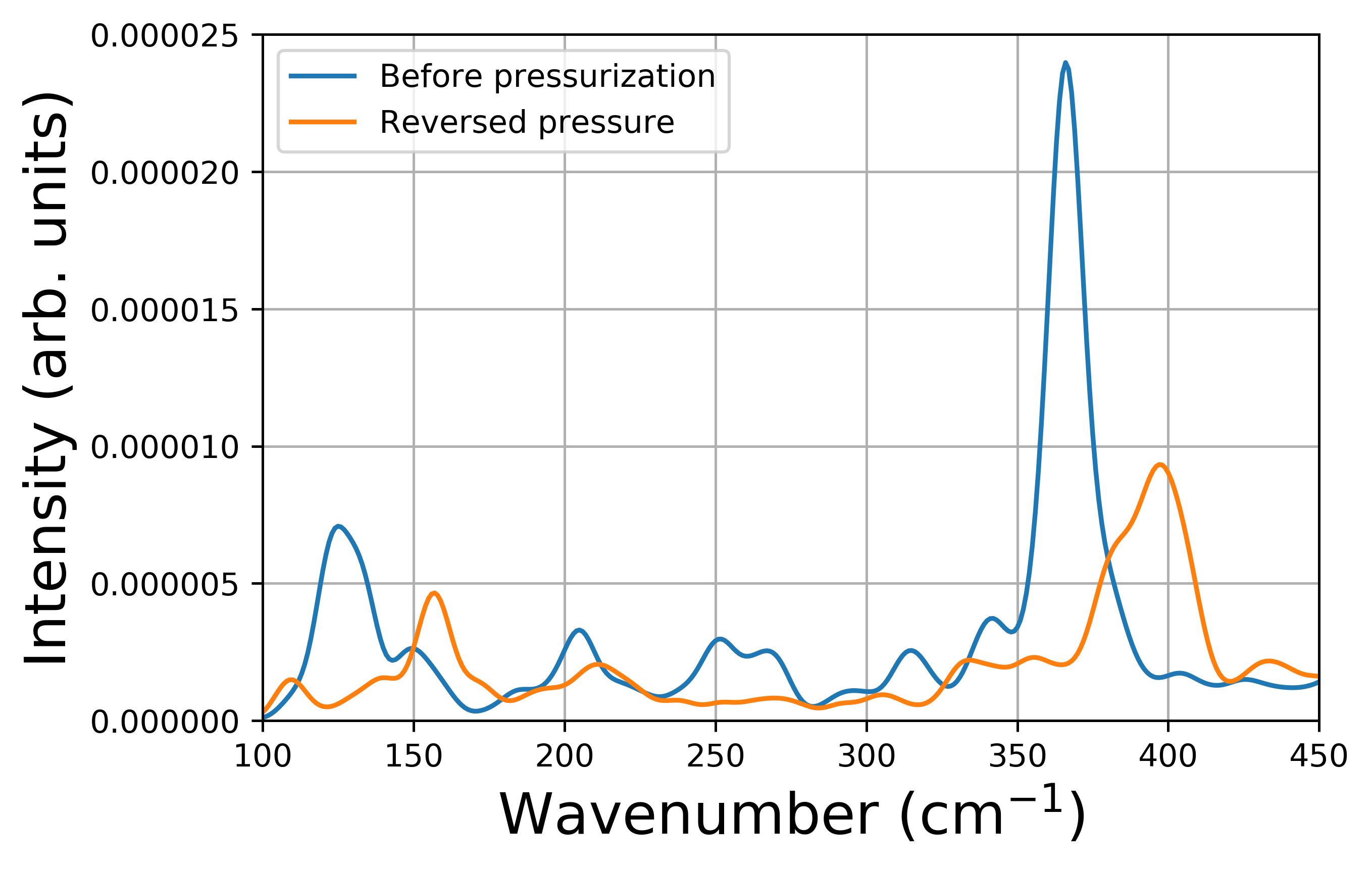}\\
	\includegraphics[width=0.45\textwidth]{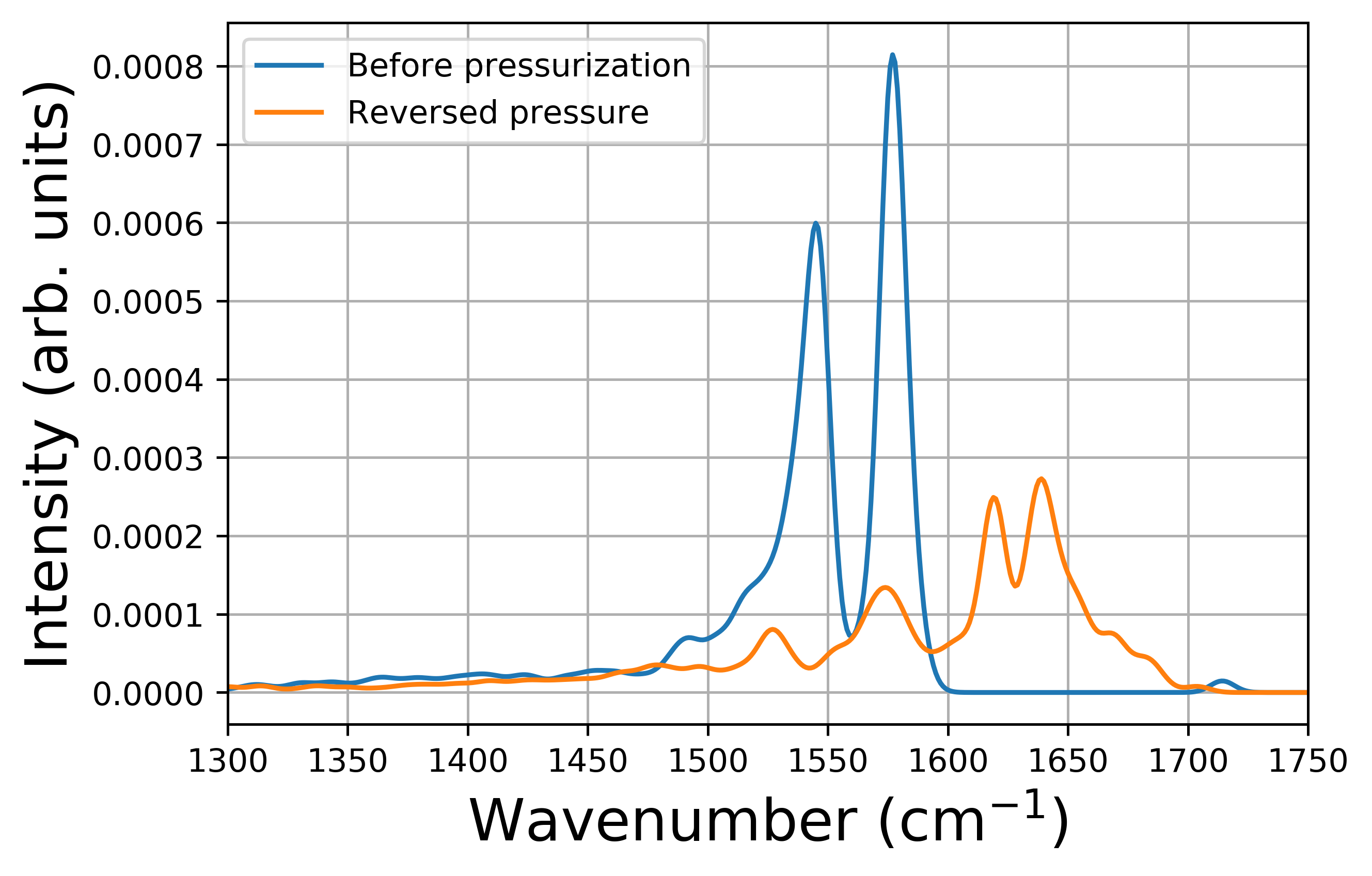}
	\caption{Comparison of the Raman spectrum of the defect CNT bundle before and after pressurization.}
	\label{fig:recovered_spectrum}
\end{figure}
Fig.~\ref{fig:recovered_spectrum} shows the Raman spectrum of the CNTs before and after pressurization. After the pressure was reversed, the intensity of the Raman active modes decreased significantly in the low and high frequency regions. The spectrum obtained after pressurization is also blue-shifted. When pristine CNTs are subjected to pressure, the Raman active modes decrease intensity and become blue-shifted (see Fig.~\ref{fig:raman_pressure_pristine}). Hence the observed intensity decrease and blue-shift after the pressure is removed for the defect nanotube indicate that the CNT bundle does not go back to the original state, as is already clear from the geometry in Fig.~\ref{fig:recovered}.\\ 
In the low frequency region of the spectrum it is observed that the radius dependent modes such as the E$_{2g}$ around \SI{155}{cm^{-1}} and the A$_{1g}$ (RBM) around \SI{400}{cm^{-1}} are still present. This signature indicates that the circular geometry of the tubes is still present though distorted. The RBM of the recovered nanotubes is broader compared to the initial nanotubes, this has been attributed to the enhanced tube interaction in the literature \cite{loa2003raman,hearne2021effects}. As both defects and interlinks influence the breathing motion, either can be attributed. The broadening of the RBM similarly indicates the existence of different regions within the CNTs, with more or less defects/interlinks.\\
Like the low frequency modes, a significant intensity drop and shift is observed in the high frequency region of the Raman spectrum shown in Fig.~\ref{fig:recovered_spectrum}, indicating that the CNTs do not return to their initial state. The D to $G^+$ ratio was computed as an indication for the defect concentration (or damage) of the CNT bundle. Before pressurization, the value of the D to G$^+$ ratio was found to be $\sim 0.019$, whereas after pressurization the value was found to be $ \sim 0.067$. The higher ratio could be attributed to the sp$^3$ interlinks, as suggested by \cite{hearne2021effects}, however, the vacancy reconstruction discussed in Section~\ref{sec:vac_recon} could contribute to the increased ratio. As a result, the increased ratio should not be ascribed only to the existence of sp$^3$ interlinks.

\section{Conclusions}
Molecular Dynamics simulations show that the introduction of vacancy defects reduces the collapse pressure of CNTs as is expected and that interlinks are likely to form on the high curvature points of the collapsed CNTs, which is consistent with previous works \cite{jensen2020toward}. Pristine CNTs fail to induce any interlinking at low temperatures. At elevated temperatures interlinks are observed, however the CNTs get damaged. Stone-Wales defects exhibit similar behavior to pristine CNTs in that, at low temperatures, they do not induce any CNT-CNT interlinks. Vacancy defects tend to reconstruct and form different kinds of topological defects under pressure. However, increasing the number of vacancy defects on the CNT yields positive results. Interlinks were observed after the pressure was removed.\\
Using DFT the Raman spectrum of some CNT systems studied using MD were computed. Under pressure a blue-shift of the Raman spectrum of pristine CNTs was observed along with along with a decrease in intensity. This was observed for all prominent Raman modes except for the low frequency E$_{1g}$ mode. The introduction of defects causes multiple modes to become Raman active. The most notable is the D-band, which is expected to arise in defect systems \cite{skakalova2007intermediate, hulman2005raman}. One crucial observation from the recovered spectrum was the increased D to G$^+$ ratio. In some studies, this ratio has been interpreted as an indicator for sp$^3$ interlinks \cite{hearne2021effects}, which is not supported by the results in this study.

\section{Acknowledgements}
The authors would like to thank the DSI-NRF Centre of Excellence in Strong Materials (CoE-SM) for support.


\bibliography{mybibfile}

\end{document}